\documentclass[aps,prb,twocolumn,superscriptaddress,showpacs,floatfix]{revtex4-1}

\usepackage{times}
\usepackage{graphicx}
\usepackage{dcolumn}
\usepackage{bm}
\usepackage{color}
\usepackage{longtable}
\usepackage{booktabs}
\usepackage{amssymb}
\usepackage{textcomp}
\usepackage{upgreek}

\begin{document}

\title{Superconducting gap evolution in overdoped BaFe$\mathbf{_{2}}$(As$\mathbf{_{1-x}}$P$\mathbf{_x}$)$\mathbf{_2}$ single crystals through nanocalorimetry}

\author{D. Campanini}
\affiliation{Department of Physics, Stockholm University, SE-106 91 Stockholm, Sweden}

\author{Z. Diao}
\affiliation{Department of Physics, Stockholm University, SE-106 91 Stockholm, Sweden}

\author{L. Fang}
\affiliation{Materials Science Division, Argonne National Laboratory, 9700 South Cass Avenue, IL 60439, USA}

\author{W-K. Kwok}
\affiliation{Materials Science Division, Argonne National Laboratory, 9700 South Cass Avenue, IL 60439, USA}

\author{U. Welp}
\affiliation{Materials Science Division, Argonne National Laboratory, 9700 South Cass Avenue, IL 60439, USA}

\author{A. Rydh}\email{andreas.rydh@fysik.su.se}
\affiliation{Department of Physics, Stockholm University, SE-106 91 Stockholm, Sweden}

\date{\today}

\begin{abstract}
We report on specific heat measurements on clean overdoped $\mathrm{BaFe_{2}(As_{1-x}P_x)_2}$ single crystals performed with a high resolution membrane-based nanocalorimeter. A nonzero residual electronic specific heat coefficient at zero temperature $\gamma_\mathrm{r}={C/T}|_{T \to 0}$ is seen for all doping compositions, indicating a considerable fraction of the Fermi surface ungapped or having very deep minima. The remaining superconducting electronic specific heat is analyzed through a two-band s-wave $\alpha$ model in order to investigate the gap structure. Close to optimal doping we detect a single zero-temperature gap of $\Delta_0 \sim 5.3\,\mathrm{meV}$, corresponding to $\Delta_0 / k_\mathrm{B} T_\mathrm{c} \sim 2.2$. Increasing the phosphorus concentration $x$, the main gap reduces till a value of $\Delta_0 \sim 1.9\,\mathrm{meV}$ for $x = 0.55$ and a second weaker gap becomes evident. From the magnetic field effect on $\gamma_\mathrm{r}$, all samples however show similar behavior [$\gamma_\mathrm{r}(H) - \gamma_\mathrm{r}(H=0) \propto H^n$, with $n$ between 0.6 and 0.7]. This indicates that, despite a considerable redistribution of the gap weights, the total degree of gap anisotropy does not change drastically with doping.
\end{abstract}


\maketitle

\section{Introduction}
The symmetry and structure of the superconducting gap in iron-based superconductors is a highly debated topic \cite{Hirschfeld2011}. In contrast with cuprates, where a \emph{d}-wave symmetry is predominant \cite{Tsuei2000, Hashimoto2014}, iron-based superconductors present a relatively wide range of possible scenarios. The leading hypothesis is a multi-band $s_{\pm}$ symmetry \cite{Mazin2009, Mazin2010, Hirschfeld2011}, where the order parameter assumes opposite sign on different sheets of the Fermi surface, but remains relatively constant in amplitude along them (no nodes). Despite the fact that many iron-based superconductors present a gap structure compatible with a nodeless $s_{\pm}$, nodal behavior has been observed in several compounds \cite{Grafe2008,Gordon2009,Reid2010,Kim2012,Murphy2013,Hashimoto2010,Nakai2010,Yamashita2011,Zhang2012,Yoshida2014}. \emph{Accidental} nodes, not due to the gap symmetry but to strong variations of the gap amplitude along a Fermi surface sheet, have then been taken into account in theoretical models to reconcile the apparent $s_{\pm}$ symmetry with zeroes in the superconducting gap \cite{Kuroki2009,Thomale2011}. A clear picture has however not been achieved yet and more experimental and theoretical efforts are required.
The isovalently doped  system BaFe$_2$(As$_{1-x}$P$_x$)$_2$ of the 122 family is particularly interesting as signs of nodal behavior have been detected even at optimal doping \cite{Hashimoto2010,Nakai2010,Yamashita2011,Zhang2012,Yoshida2014}. This is in stark contrast with its hole- and electron-doped counterparts which, despite showing similar phase diagrams and critical temperatures, are believed to be nodeless at least at optimal doping \cite{Mu2009,Luo2009,Nakayama2011,Tanatar2010}. Specific heat measurements have been performed on the hole-doped Ba$_{1-x}$K$_x$Fe$_2$As$_2$ \cite{Mu2009,Popovich2010,Storey2013} and on the electron-doped Ba(Fe$_{1-x}$Co$_x$)$_2$As$_2$ \cite{Hardy2010,Hardy2010PRB,Gofryk2010}. While some features are common to both compounds, e.g., multi-gap behavior and strong-coupling values of the main gap amplitude, others are substantially different, such as the values of the residual specific heat coefficient $\gamma_\mathrm{r}$ and the magnetic field- and doping dependence of the specific heat.

In this work, we study the specific heat of BaFe$_2$(As$_{1-x}$P$_x$)$_2$ single crystals in the overdoped regime.
The low temperature electronic specific heat provides information about the gap amplitudes, while its magnetic field dependence at low temperatures reveals the gap anisotropy. A sizable value of the residual electronic specific heat coefficient $\gamma_\mathrm{r}$, amounting to $17\/\%$ near optimum doping, growing to $\sim 30\/\%$ with doping, shows that a considerable part of the Fermi surface is ungapped or presents broad deep minima. The remaining superconducting specific heat is fitted to a two-band $\alpha$-model, which is found to represent the experimental data well. Close to optimal doping, the best fit is obtained with a single gap function with a zero-temperature energy gap $\Delta_0 \sim 5.3\,\mathrm{meV}$. However, a second gap becomes evident as $x$ increases. The magnetic field dependence of the zero temperature specific heat coefficient reveals a sublinear behavior [$\Delta\gamma = \gamma_\mathrm{r}(H)-\gamma_\mathrm{r}(H=0) \propto H^{(0.6-0.7)}$], largely independent of doping.

\section{Experimental details}
\begin{figure}[t]
	\includegraphics[width=\linewidth]{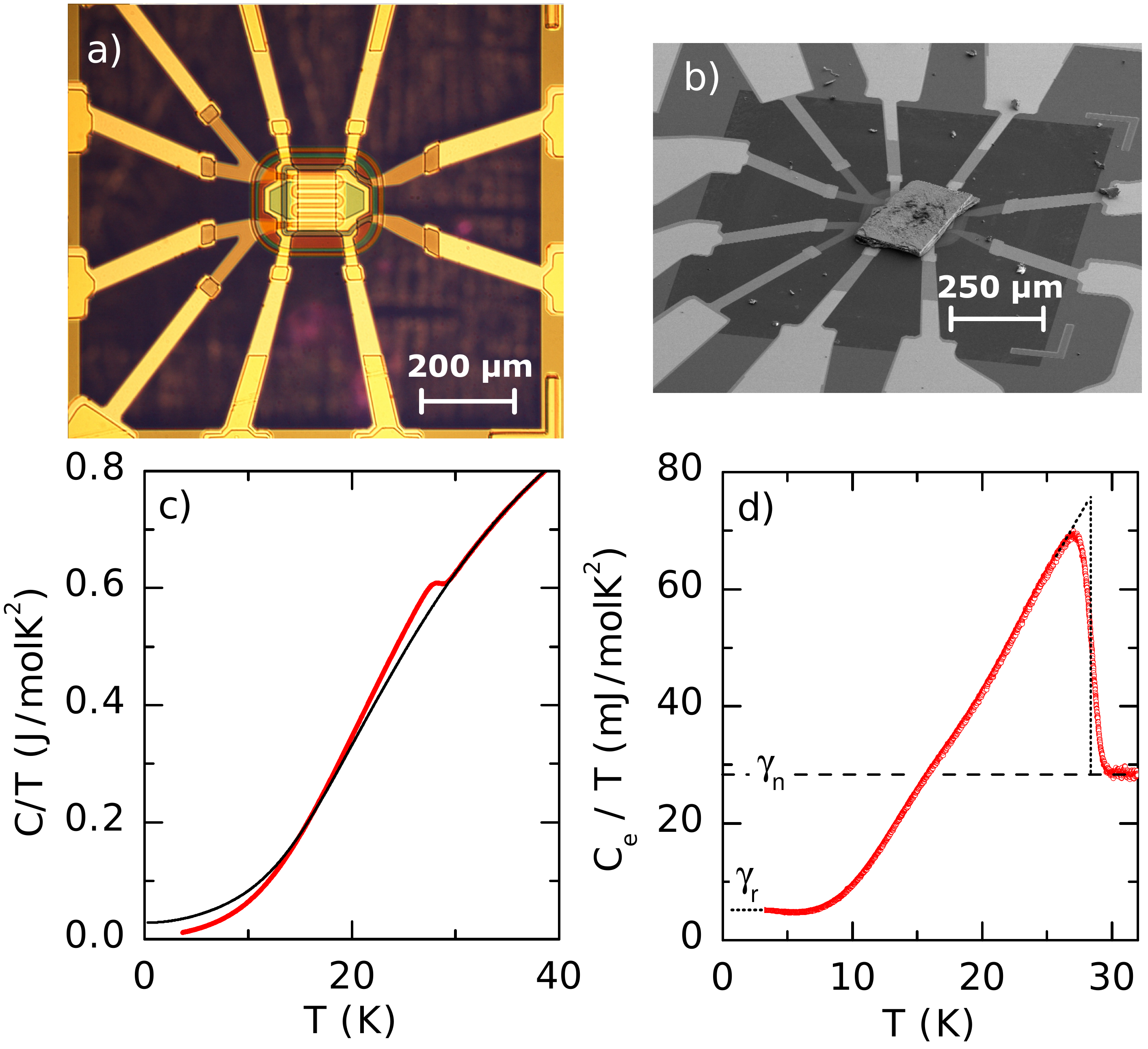}
	\caption{(a) Optical microscope image of the sample side of a typical nanocalorimeter. (b) SEM image of the sample cell, with the most overdoped sample ($T_\mathrm{c} = 12.5\,$K) attached to it. (c) Specific heat plotted as $C/T$ of the $T_\mathrm{c}=28.4\,\mathrm{K}$ sample as a function of temperature $T$. Red circles represent the experimental data, while the black curve is an interpolation of the normal state signal with a Debye-Sommerfeld function. (d) Temperature dependence of the electronic specific heat $C_\mathrm{e}$ at low temperatures shown as $C_\mathrm{e}/T$ for the same sample as in (c).}
	\label{FigMeasurement}
\end{figure}

The high-purity $\mathrm{BaFe_2(As_{1-x}P_x)_2}$ single crystals investigated in this work were grown with a self-flux method \cite{Chaparro2012}. Three crystals with composition $x = 0.32, 0.50$ and $0.55$ were selected and cleaved in order to obtain plate-like samples of side $\sim 100-200\,\upmu\mathrm{m}$. Optical microscopy observation comfirmed that all samples had shiny surfaces with no identifiable secondary phase inclusions. $x = 0.32$ corresponds to near optimal doping with $T_\mathrm{c} = 28.4\,\mathrm{K}$, while $x = 0.50$ and $x = 0.55$ are in the overdoped regime with $T_\mathrm{c} = 18.2\,\mathrm{K}$ and $12.5\,\mathrm{K}$, respectively. Specific heat was measured with a differential membrane-based nanocalorimeter applying an AC-method with phase stabilized frequency feedback \cite{Tagliati2011, Tagliati2012}. The sample side cell of a typical calorimeter device is illustrated in Fig.~\ref{FigMeasurement}(a). The active area is a stack of thin films in the center of a $\mathrm{Si_3N_4}$ membrane, whose key elements are a GeAu thermometer and Ti (AC and DC) heaters. A minute amount of Apiezon-N grease was used to attach the crystals to the calorimeter, as shown in Fig.~\ref{FigMeasurement}(b). The grease specific heat was measured separately in order to subsequently extract the intrinsic sample response.

\section{Results and discussion}
The temperature dependence of the specific heat plotted as $C/T$ of the $T_\mathrm{c}=28.4\,\mathrm{K}$ crystal is reported in Fig.~\ref{FigMeasurement}(c). The transition to the superconducting state is apparent as a small peak at $T_\mathrm{c}$. The specific heat above the transition is fitted with a Debye-Sommerfeld function and extended below the transition with the requirement of entropy conservation at $T_\mathrm{c}$. The electronic specific heat is obtained by subtracting the phonon contribution from the total specific heat and is shown as $C_\mathrm{e}/T$ in Fig.~\ref{FigMeasurement}(d) for the same sample. A similar procedure was applied for the $T_\mathrm{c}=18.2\,\mathrm{K}$ and $12.5\,\mathrm{K}$ samples. In the latter case, with a relatively low upper critical field $H_{c2}$, it is possible to verify that the normal state when applying a magnetic field of 5\,T actually corresponds to the one calculated with the Debye-Sommerfeld function.

In all studied samples, a sizable residual specific heat $\gamma_\mathrm{r}$ is found, ranging from $17\/\%$ of the normal state coefficient $\gamma_\mathrm{n}$ for $T_\mathrm{c}=28.4\,\mathrm{K}$ to 31 and $28\/\%$ for the $T_\mathrm{c}=18.2\,\mathrm{K}$ and $T_\mathrm{c}=12.5\,\mathrm{K}$ samples, respectively. This term is due to the presence of non-superconducting quasiparticles. Since the crystals are nicely shaped with freshly cleaved surfaces, macroscopic secondary phases are unlikely to be the cause of this high $\gamma_\mathrm{r}$. Therefore, we believe that the residual term is due to a part of the Fermi surface being ungapped or presenting deep broad minima in the order parameter suppressed by weak disorder. Values of $\gamma_\mathrm{r}/\gamma_\mathrm{n}$ in the range of $10-20\/\%$  have been reported for Co-doped samples close to optimal doping \cite{Hardy2010PRB,Hardy2010} ($\gamma_\mathrm{r}/\gamma_\mathrm{n} \sim 5\/\%$ at optimal doping), much higher than for K-doped samples, where $\gamma_\mathrm{r}/\gamma_\mathrm{n}$ accounts to only a few percent \cite{Popovich2010,Storey2013}. This is in agreement with a fully gapped state detected in K-doped samples over a wide range of the phase diagram \cite{Nakayama2011}, while nodal behavior is detected from thermal conductivity measurements in Co-doped samples as soon as $x$ moves away from optimal doping \cite{Reid2010}. The rather high value of $\gamma_\mathrm{r}/\gamma_\mathrm{n}$ in our samples is in agreement with nodes/minima measured from angle-resolved photoemission spectroscopy (ARPES) \cite{Zhang2012, Yoshida2014}, penetration depth \cite{Hashimoto2010}, thermal conductivity \cite{Hashimoto2010,Yamashita2011} and nuclear magnetic resonance \cite{Nakai2010}. 
The absolute value of $\gamma_\mathrm{r}$ in our samples does not show an increase with doping as reported for Co-doped Ba-122 \cite{Hardy2010}. Since quantum oscillations are observed only in overdoped $\mathrm{BaFe_2(As_{1-x}P_x)_2}$ samples \cite{Shishido2010, Analytis2010}, it is likely that phosphorus doping decreases the disorder in the crystal, avoiding a strong increase in $\gamma_\mathrm{r}$.

\begin{figure}[b]
	\includegraphics[width=1\linewidth]{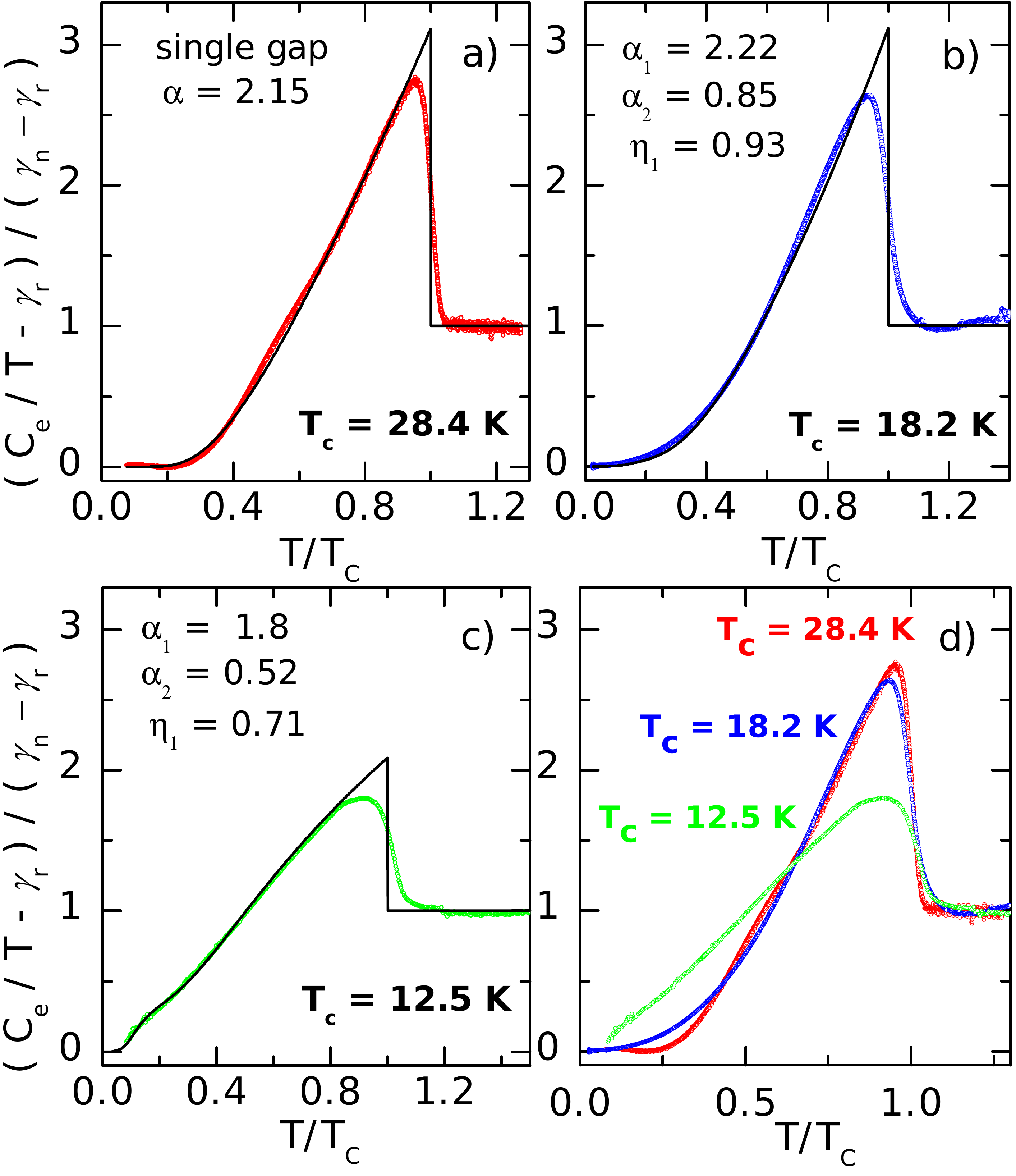}
	\caption{Temperature dependence of the reduced electronic specific heat $C_\mathrm{e}/T$, subtracted by $\gamma_\mathrm{r}$ and normalized by $(\gamma_\mathrm{n}-\gamma_\mathrm{r})$. Data points are represented by colored dots, the fits according to a two-gap $\alpha$-model by black lines. The fitting parameters are reported next to each curve. (a) $T_\mathrm{c}=28.4\,\mathrm{K}$. (b) $T_\mathrm{c}=18.2\,\mathrm{K}$. (c) $T_\mathrm{c}=12.5\,\mathrm{K}$. (d) Resume of the 3 data sets.}
	\label{FigAlphaModel}
\end{figure}

In order to obtain the superconducting contribution $C_\mathrm{es}/T$ the residual term $\gamma_\mathrm{r}$ is subtracted from the total $C/T$. The result is shown in Fig.~\ref{FigAlphaModel} for the three doping levels analyzed. The data are presented as a function of $T/T_\mathrm{c}$ and normalized by ($\gamma_\mathrm{n}-\gamma_\mathrm{r}$) in order to allow a direct comparison between them. $C_\mathrm{es}/T$ is then fitted to a two-band implementation of the phenomenological $\alpha$-model \cite{Bouquet2001,Johnston2013}, which assumes full s-wave gaps both closing at $T_\mathrm{c}$ and a BCS temperature dependence. ARPES measurements indicate that this is indeed a reasonably good assumption \cite{Zhang2012, Yoshida2014} and as a consequence the $\alpha$-model is able to give a good description of the specific heat data. An alternative approach would be the $\gamma$-model of Kogan, Martin and Prozorov \cite{Kogan2009}, which can be used for a general temperature dependence of the gap, but which requires knowledge about the Fermi velocity and density of states on the different bands. 
The electronic specific heat according to the $\alpha$-model is calculated as \cite{Johnston2013}:
\begin{equation}
\frac{C_\mathrm{es}(t)}{(\gamma_\mathrm{n} - \gamma_\mathrm{r})T_\mathrm{c}} = \frac{6\alpha^3}{\pi^2t}\int_{0}^{\infty}f(1-f)\left(\frac{\tilde{E}^2}{t}-\frac{1}{2}\frac{\mathrm{d}\tilde{\Delta}^2}{\mathrm{d}t}\right)\mathrm{d}\tilde{\epsilon},
\end{equation}
where  $\alpha = \Delta_0/k_\mathrm{B}T_\mathrm{c}$ is an adjustable parameter, $t = T/T_\mathrm{c}$, $\tilde{\epsilon} = \epsilon/\Delta_0$ is the normalized single-particle energy, $\tilde{E}=\sqrt{\tilde{\epsilon}^2 + \tilde{\Delta}^2}$ is the normalized energy of elementary quasiparticle excitations, $f(\tilde{E},t,\alpha)=\left[{\exp{(\alpha \tilde{E}/t)}+1}\right]^{-1}$ is the Fermi-Dirac distribution function, and $\tilde{\Delta}(t)=\Delta(t)/\Delta_0$ is the order parameter normalized by itz zero temperature value.  The total electronic specific heat $C_\mathrm{es}$ is then considered as the sum of two independent contributions $C_1$ and $C_2$, given by two different $\alpha$ values, $\alpha_1$ and $\alpha_2$. A weight $\eta_1$ is associated with $C_1$ and $\eta_2 = 1-\eta_1$ with $C_2$. A routine was implemented in order to vary the three free parameters $\alpha_1$, $\alpha_2$ and $\eta_1$ to minimize the root-mean-square deviation between the data and the model function.

\begin{table}
\caption{Superconducting parameters for $\mathrm{BaFe_2(As_{1-x}P_x)_2}$ obtained from electronic specific heat.}
\begin{tabular*}{\linewidth}{@{\extracolsep{\fill}}lcccc}
	\hline
	\hline
	&  & \multicolumn{3}{c}{$T_\mathrm{c}$ (K)}\\
	\cline{3-5}
	Property  & Unit & 28.4 & 18.2 & 12.5\\
	\hline\\[-2.2ex]
	$\gamma_\mathrm{n}$ & $\mathrm{mJ/molK^2}$ & 28.0 & 20.2 & 19.2\\
	$\gamma_\mathrm{r} / \gamma_\mathrm{n}$ & \% & 17 & 31 & 28\\
	$\alpha_{1}$  & & 2.15 & 2.22 & 1.8\\
	$\alpha_{2}$  & & - & 0.85 & 0.52\\
	$\Delta_{1}$ & meV & 5.3 & 3.5 & 1.9\\
	$\Delta_{2}$ & meV	& - &	1.3 & 0.6\\
\footnote{Weight associated with the first gap}$\eta_1$ & \% & 100 & 93 & 71\\
\footnote{Average gap obtained according to Eq.~(\ref{eq:average})}$\Delta_\mathrm{avg}$ & meV & 5.3 & 3.4 & 1.7\\
\footnote{Average gap obtained according to Eq.~(\ref{eq:CEBCS})}$\Delta_\mathrm{cond}$ & meV & 4.6 & 2.9 & 1.7\\
\footnote{From fittings of $\Delta\gamma(H)$ to the function $\Delta\gamma = A\cdot(\mu_0H)^{n}$}$n$ & & $0.64(6)$ & $0.66(1)$ & $0.68(2)$\\
	\hline
	\hline
\end{tabular*}
\label{tab:SCParameters}
\end{table}

The best fitting parameters obtained are presented in Fig.~\ref{FigAlphaModel}(a)-(c) for each curve and summarized in Table~\ref{tab:SCParameters}. Their uncertainties are estimated to $\sim 5\/\%$. All fitting curves allow a fairly good representation of the data.
For $T_\mathrm{c}=28.4\,\mathrm{K}$ the experimental curve clearly saturates at low temperatures ($T/T_\mathrm{c} < 0.25$), leading to the conclusion that no minor gap should contribute, at least within our measurement resolution. The best fit is in fact given by a single gap function, with $\alpha = 2.15$ ($\Delta = 5.3\,\mathrm{meV}$). For $T_\mathrm{c}=18.2\,\mathrm{K}$, the curve no longer saturates at low temperatures and a contribution from a second smaller gap has to be considered. The fitting routine gives a main gap $\alpha_1 = 2.22$ ($\Delta_{1} = 3.5\,\mathrm{meV}$), with a considerable weight $\eta_1 = 93\/\%$ and a minor $\alpha_2 = 0.85$ ($\Delta_{2} = 1.3\,\mathrm{meV}$) for the remaining $\eta_2 = 7\/\%$. For $T_\mathrm{c}=12.5\,\mathrm{K}$, the main peak reduces considerably in amplitude ($\alpha_1 = 1.8$, with an associated $\Delta_{1} = 1.9\,\mathrm{meV}$). Moreover, the second gap contribution is much more pronounced ($\eta_2 = 29\/\%$). Its amplitude is $\Delta_{2} = 0.6\,\mathrm{meV}$ ($\alpha_2 = 0.52$).
The $\alpha_1$s associated with the main gap are all higher than the BCS value $\alpha_\mathrm{BCS} = 1.764$, indicating strong-coupling. However, the $T_\mathrm{c}=12.5\,\mathrm{K}$ sample presents an $\alpha_1$ much closer to $\alpha_\mathrm{BCS}$ than the other two samples.
A weighted average of the gap amplitudes $\Delta_\mathrm{avg}$ is calculated from the $\alpha$-model results for the three doping compositions:
\begin{equation}
\label{eq:average}
\Delta_\mathrm{avg} = \sqrt{{\Delta_1}^2\cdot\eta_1+{\Delta_2}^2\cdot\eta_2}
\end{equation}
The results are listed in Table~\ref{tab:SCParameters}. The average value is dominated by the main gap in all samples.
The obtained $\Delta_\mathrm{avg}$s are compared to the gap amplitudes calculated from the BCS-style relation for the condensation energy
\begin{figure}[b]
	\includegraphics[width=0.53\linewidth]{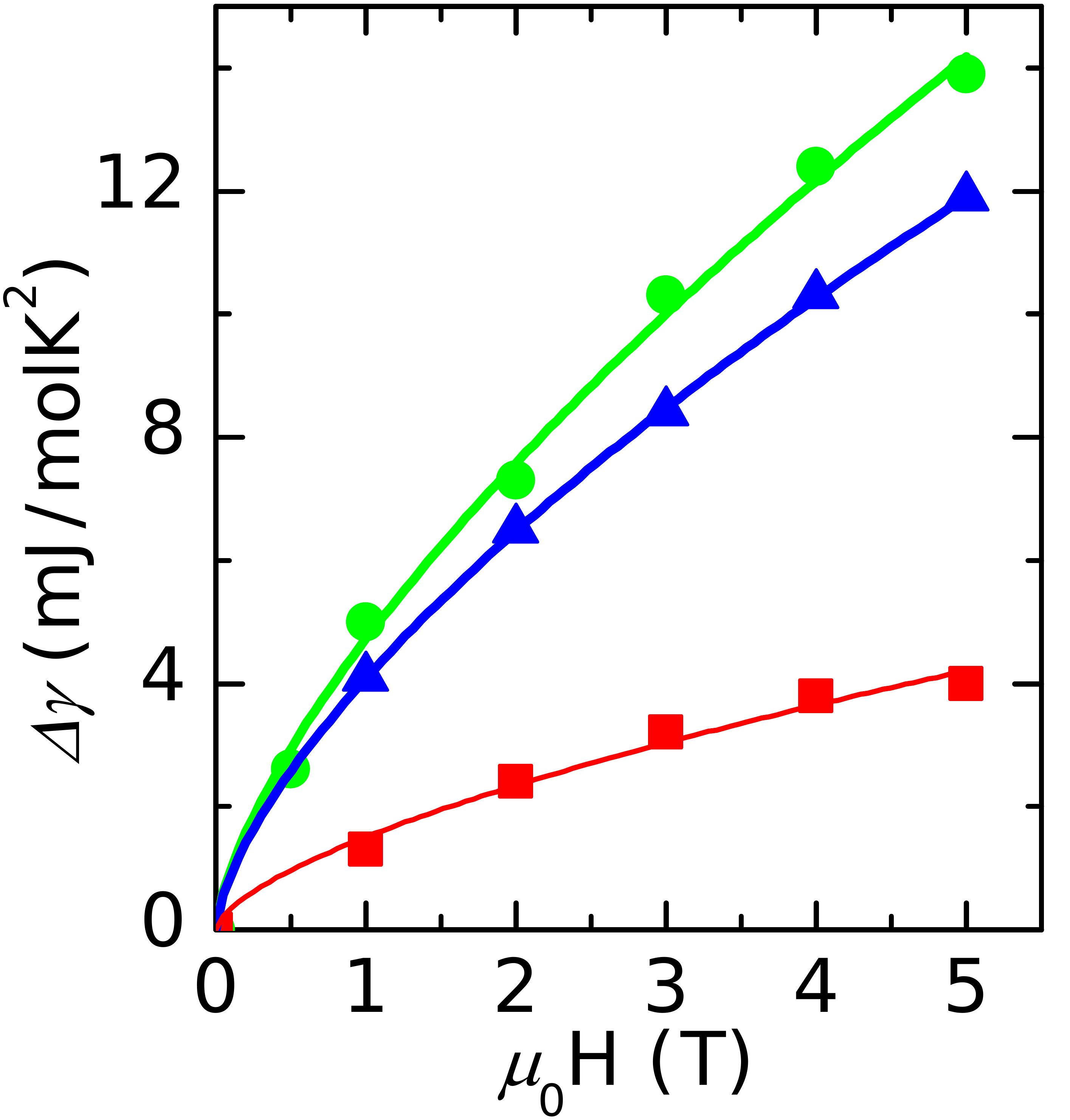}
	\caption{Field dependence of $\Delta\gamma = \gamma_\mathrm{r}(H) - \gamma_\mathrm{r}(H=0)$ for $T_\mathrm{c}=28.4\,\mathrm{K}$ (red squares), $T_\mathrm{c}=18.2\,\mathrm{K}$ (blue triangles) and $T_\mathrm{c}=12.5\,\mathrm{K}$ (green circles). The curves are fits of the type $\Delta\gamma = A\cdot(\mu_0H)^{n}$, with $n$ given in Table~\ref{tab:SCParameters} for the different samples.}
	\label{FigFieldDependence}
\end{figure}
\begin{equation}
\label{eq:CEBCS}
\Delta F = \frac{(\gamma_\mathrm{n}-\gamma_\mathrm{r})}{\gamma_\mathrm{n}}N(E_F)\Delta_\mathrm{cond}^2/4,
\end{equation}
where $N(E_F)$ is the density of states at the Fermi energy, $N(E_F) = 3\gamma_\mathrm{n}/\pi^2k_\mathrm{B}^2$. $\Delta F$ is obtained by integration of the $\Delta C (T)$ curve between 0 and $T_\mathrm{c}$ \cite{Diao2014}, where $\Delta C$ is the difference in specific heat between the superconducting and the normal state. The gap values obtained from $\Delta F$ are reported in Table~\ref{tab:SCParameters} as well. The most overdoped sample, which has an energy gap ratio very close to the BCS value, shows very good agreement between $\Delta_\mathrm{avg}$ and $\Delta_\mathrm{cond}$. For the other two samples, the average gaps are around $15\/\%$ higher than $\Delta_\mathrm{cond}$. This behavior is likely due to strong-coupling at these dopings, for which Eq.~(\ref{eq:CEBCS}), valid under the assumption of weak-coupling, underestimates $\Delta$. It is in fact theoretically expected that strong-coupling superconductors show a condensation energy lower than that expected in the BCS case \cite{Haslinger2003, Carbotte1990}. A lower $\Delta F$ then corresponds to a lower $\Delta_\mathrm{cond}$. The $\alpha$-model instead phenomenologically takes into account strong-coupling effects and gives a better representation of the experimental data.

The gap amplitude close to optimal doping, $\Delta_\mathrm{avg} = 5.3\,\mathrm{meV}$, can be compared with the ARPES values of Refs. ~[\onlinecite{Zhang2012}] and [\onlinecite{Yoshida2014}], which report $\sim5-8\,\mathrm{meV}$ for the gap on the hole pockets and $\sim7-8\,\mathrm{meV}$ on the electron pockets. The values seem in good agreement considering that we are slightly out of optimally doped conditions and that the specific heat signal is more sensitive to the hole pockets (see [\onlinecite{Kim2010}] and references therein). P overdoping has the effect of reducing the main superconducting gap and increasing the weight of the second gap. In order to better visualize the evolution of the gap with doping, experimental data for the three compositions studied are plotted together in Fig.~\ref{FigAlphaModel}(d). The behavior is different in comparison with Co-doped samples \cite{Gofryk2010,Hardy2010}, where the two gaps have fairly constant weights throughout the entire doping range. It is instead in agreement with K-doped samples \cite{Storey2013}, where the weight of the second gap is increasing with doping, as in the present case.

The field dependence of the zero temperature specific heat coefficient $\gamma_\mathrm{r}$ is analyzed in order to extract information on the gap anisotropy. Specific heat was measured in magnetic fields up to 5\,T for all samples. 5\,T corresponds to 11\/\%, 39\/\%, and 93\/\% of $H_{c2}$ for the samples with $T_\mathrm{c}=28.4\,\mathrm{K}, 18.2\,\mathrm{K}$, and $12.5\,\mathrm{K}$, respectively. The zero temperature $\gamma_\mathrm{r}$ values are obtained by linear extrapolations of the low-temperature $C_\mathrm{e}/T$ curves. The resulting $\gamma_\mathrm{r}(H)$ values are shown in Fig.~\ref{FigFieldDependence} as a function of applied field $\mu_{0}H$ for the three doping compositions. The curves are vertically shifted down to zero by subtracting the zero field $\gamma_\mathrm{r}(H=0)$ from all $\gamma_\mathrm{r}(H)$ values.
A nodeless s-wave order parameter is expected to give a linear dependence of $\gamma_\mathrm{r}$ as a function of field \cite{Caroli1964}, while a d-wave order parameter a square root dependence with field \cite{Volovik1993}. An intermediate behavior is generally interpreted in iron-based superconductors as due to the presence of gaps with different amplitudes \cite{Gofryk2010,Storey2013}. The gap imbalance becomes more pronounced as $\gamma_\mathrm{r}(H)$ deviates further from a linear relation. All curves were fitted with a function of the type $\Delta\gamma = A\cdot(\mu_0H)^n$, with $A$ and $n$ being the fitting parameters. The obtained exponents $n$ are reported in Table~\ref{tab:SCParameters}. The exponent is in the range of $0.6$ to $0.7$ for all three samples. Previous measurements at optimal doping \cite{Wang2011} show as well a very similar field dependence for fields up to 4\,T, while a crossover to a linear behavior is found at higher fields. This behavior is interpreted in terms of a double gap system, in which the Volovik-like trend at low fields is due to a strongly anisotropic gap, while the linear component at high fields is due to a second isotropic gap. Even if we see no sign of a second gap from the temperature dependence of the specific heat of the $T_\mathrm{c} = 28.4\,\mathrm{K}$ sample as shown in Fig.~\ref{FigAlphaModel}(a), such a gap should be expected in order to explain the field dependence of $\Delta\gamma$. Its weight is however at most about 5\/\%, given by the uncertainty of the fitting parameters.
The exponent $n$ displays no clear trend with doping. This shows that, despite different weights of the gaps along the doping range, the total degree of anisotropy, as measured by the field dependence of $\Delta\gamma$, stays constant. This is in qualitative agreement with the theory of Bang \cite{Bang2010}, which shows that for a two-band $\mathrm{s_{\pm}}$ state in the presence of impurity scattering the field dependence of $\Delta\gamma$ mainly depends on the ratio between the two gap amplitudes $\Delta_2/\Delta_1$ and not on their weights.

\section{Conclusions}
In conclusion, from high-resolution specific heat measurements on overdoped $\mathrm{BaFe_2(As_{1-x}P_x)_2}$ single-crystals we observe: (i) A sizable residual term $\gamma_\mathrm{r}$ at all dopings, sign of a considerable part of the Fermi surface presenting no superconducting gap or very deep minima. (ii) A main gap amplitude in the strong-coupling limit ($\alpha = \Delta_0/k_\mathrm{B}T_\mathrm{c}$ between 1.8 and 2.22), in agreement with ARPES measurements near optimal doping. (iii) A reduction of the main gap weight with doping, which passes from dominating the specific heat signal near optimal doping to a $71\/\%$ of the total weight for $T_\mathrm{c}=12.5\,\mathrm{K}$. (iv) A high gap anisotropy, constant at all dopings.

\emph{Note:} During preparation of this paper, we became aware of a recent specific heat study \cite{Malone2014}, which shows that a single-band anisotropic gap fit gives a good representation of the experimental data as well.

\begin{acknowledgments}
This work was supported by the U.S. Department of Energy, Office of Science, Basic Energy Sciences, Materials Sciences and Engineering Division. We thank D. Nkulikiyimfura for assistance in the nanocalorimeter fabrication and are grateful for equipment supported by the K.\&A. Wallenberg foundation.
\end{acknowledgments}

\end{document}